# Astronomical Creation of Cyclic-C3H2 and Chain-C3 Due to Interstellar Deep Photoionization

Norio Ota

Graduate School of Pure and Applied Sciences, University of Tsukuba,
1-1-1 Tenoudai Tsukuba-city 305-8571, Japan;   n-otajitaku@nifty.com

Astronomical evolution mechanism of small size polycyclic aromatic hydrocarbon (PAH) was analyzed using the first principles quantum-chemical calculation. Starting model molecule was benzene ($C_6H_6$), which would be transformed to ($C_5H_5$) due to carbon defect attacked by interstellar high speed proton. In a protoplanetary disk around a young star, molecules would be illuminated by high energy photon and ionized to be cation $(C_5H_5)^{n+}$. Calculation shows that from n=0 to 3, molecule keeps original configuration. At a step of n=6, there occurs surprising creation of cyclic-$C_3H_2$, which is the smallest PAH. Astronomical cyclic-$C_3H_2$ had been identified by radio astronomy. Deep photoionization on cyclic-$C_3H_2$ brings successive molecular change. Neutral and mono-cation keep cyclic configuration. At a step of di-cation, molecule transformed to aliphatic chain-$C_3H_2$. Finally, chain-$C_3H_2$ was decomposed to pure carbon chain-$C_3$ and two hydrogen atoms. Calculated infrared spectrum of those molecules was applied to observed spectrum of Herbig Ae young stars. Observed infrared spectrum could be partially explained by small molecules. Meanwhile, excellent coincidence was obtained by applying a larger molecules as like $(C_{23}H_{12})^{2+}$ or $(C_{12}H_8)^{2+}$. Infrared observation is suitable for larger molecules and radio astronomy for smaller asymmetric molecules. It should be noted that these molecules could be identified in a natural way introducing two astronomical phenomena, that is, void-induced molecular deformation and deep photoionization.

Key words:  astrochemistry, cyclic-$C_3H_2$, chain-$C_3$, infrared astronomy, radio astronomy

## 1, INTRODUCTION

 Study on astronomical evolution mechanism of polycyclic aromatic hydrocarbon (PAH) is essential to suggest one inevitable rout of a building block of "life" in the universe. Especially, PAHs around a Herbig Ae young star is promising. Herbig Ae star is a pre-main-sequence star having a few solar masses and would be a model of our early solar system. There is an important observed infrared spectrum (IR) edited by B. Acke et al. including 53 stars (Acke et al. 2010). In our previous quantum-chemical study, we could identify these spectra by hydrocarbon pentagon-hexagon combined molecules as shown in Figure 1. The first successful identification was obtained by void induced coronene $(C_{23}H_{12})^{2+}$ (Ota 2014, 2015) applied to ubiquitously observed spectrum group (half of 53 examples). Starting molecule was coronene $(C_{24}H_{12})$ having seven carbon hexagons, which would be transformed to $(C_{23}H_{12})$ due to carbon defect created by interstellar high speed proton attack. It should be noted that there occurs Jahn-Teller quantum deformation by void creation (Jahn and Teller 1937, Ota 2018c). In a protoplanetary disk, molecules would be illuminated by high energy photon from the central star and ionized to be cation $(C_{23}H_{12})^{n+}$ (Ota 2017a, 2017b, 2017c, 2017d). Among observed infrared spectrum of Herbig Ae stars, there was another type spectrum of almost 40% of the list, which could be identified by ionized acenaphthylene $(C_{12}H_8)^{n+}$. This type was introduced starting from $(C_{13}H_9)$ having three carbon hexagons (Ota 2017e, 2018b). Through these studies, we could find two steps how to certify specific interstellar molecule, that is, (1) void-induced quantum deformation and (2) deep photo-ionization. Applying this rule, we could also find larger molecules as like $(C_{53}H_{18})^{n+}$ and $(C_{53})^{n+}$ (Ota 2018a, 2018c).  Now, question is how about small molecules applying this rule. This study will focus on finding the smallest PAH molecule starting from benzene $(C_6H_6)$. We like to report here to find cyclic-$C_3H_2$, chain-$C_3H_2$ and chain-$C_3$. Cyclic-$C_3H_2$ is well observed in interstellar dust and protoplanetary disk around a very young star by radio astronomy observations (Thaddeus et al. 1985, Matthews et al. 1985, Pety et al. 2005, Qi et al. 2013, Sakai et al. 2014, Autur de la Villarmois et al. 2018). This study will also suggest an interplay between infrared astronomy and radio astronomy.

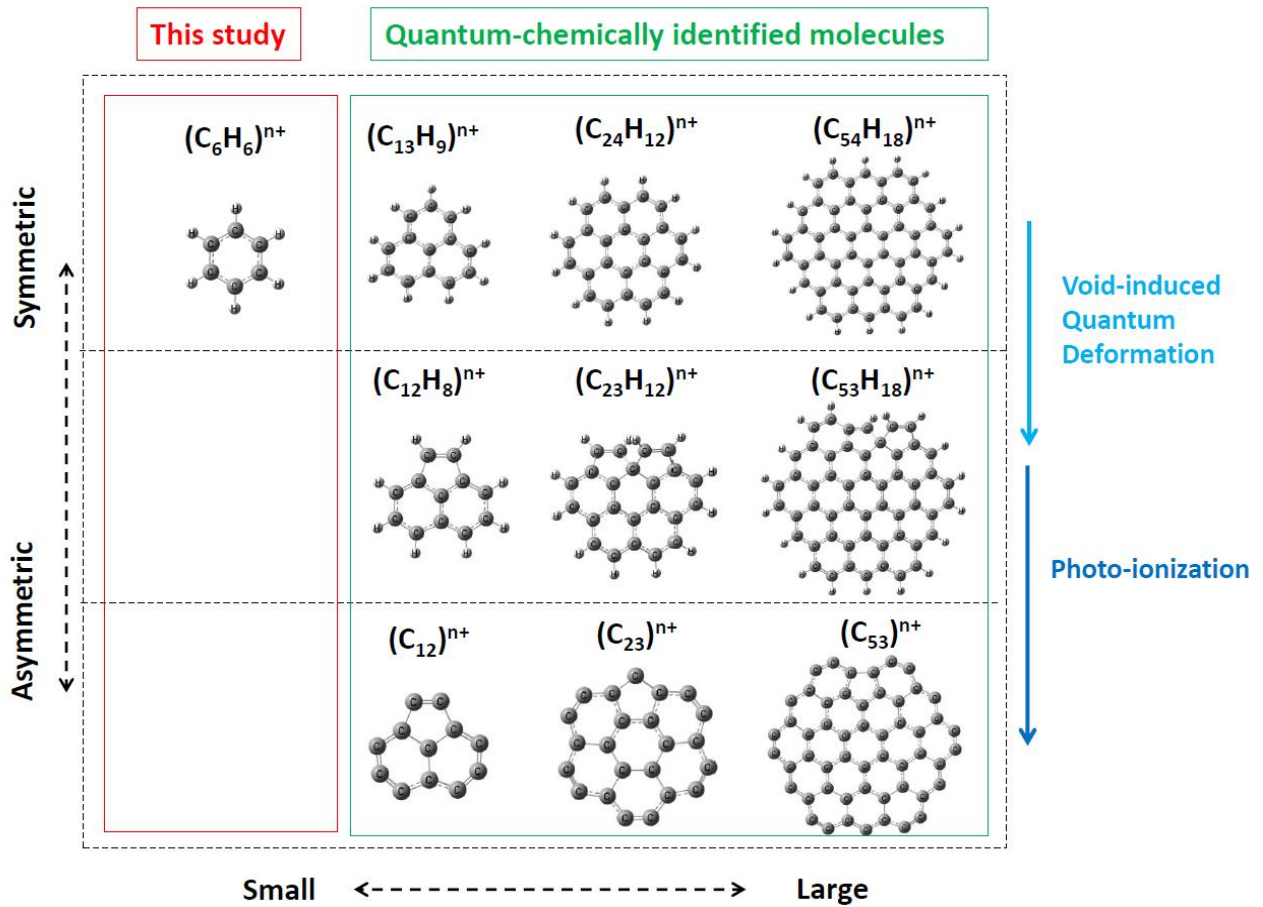

Figure 1, Quantum-chemically identified molecules applied to observed infrared spectrum of interstellar dust and Herbig Ae young stars. Those previously studied molecules were obtained by two finding rules of (1) void-induced quantum deformation and (2) deep photo-ionization. In this study, we like to find out smaller molecules starting from benzene ($C_6H_6$).

## 2, CALCULATION METHOD

In quantum chemistry calculation, we have to obtain total energy, optimized atom configuration, and infrared vibrational mode frequency and strength depend on a given initial atomic configuration, charge and spin state Sz. Density functional theory (DFT) with unrestricted B3LYP functional was applied utilizing Gaussian09 package (Frisch et al. 1984, 2009) employing an atomic orbital 6-31G basis set. The first step calculation is to obtain the self-consistent energy, optimized atomic configuration and spin density. Required convergence on the root mean square density matrix was less than $10^{-8}$ within 128 cycles. Based on such optimized results, harmonic vibrational frequency and strength was calculated. Vibration strength is obtained as molar absorption coefficient ε (km/mol.). Comparing DFT harmonic wavenumber $N_{DFT}$ (cm$^{-1}$) with experimental data, a single scale factor 0.965 was used (Ota 2015). Concerning a redshift for the anharmonic correction, in this paper we did not apply any correction to avoid over estimation in a wide wavelength representation from 2 to 30 micrometer.

Corrected wave number N is obtained simply by N (cm$^{-1}$) = $N_{DFT}$ (cm$^{-1}$) x 0.965.

Wavelength λ is obtained by λ (micrometer) = 10000/N(cm$^{-1}$).

Reproduced IR spectrum was illustrated in a figure by a decomposed Gaussian profile with full width at half maximum FWHM=4cm$^{-1}$.

3, FROM BENZENE (C6H6) TO (CYCLIC-C3H2)

In Figure 2, astrochemical evolution step started from benzene ($C_6H_6$) was illustrated. The first astronomical hypothesis is high speed proton attack on ($C_6H_6$) in interstellar space. Void induced open configuration would be suddenly transformed to closed cyclic-hydrocarbon ($C_5H_5$) by the Jahn-Teller effect (Jahn and Teller 1937). The central star may irradiate such molecules by high energy photon.  Neutral molecule ($C_5H_5$)$^{0+}$ will be transformed to mono-cation ($C_5H_5$)$^{1+}$ by 8.5eV photon excitation. Di-cation ($C_5H_5$)$^{2+}$ will be excited by 23.6eV compared with neutral one, also tri-cation ($C_5H_5$)$^{3+}$ by 46.1eV. Calculation shows that from n=0 to 3, molecule keeps original configuration. At ionization step n=4 as illustrated in Figure 3, there occurs separation to cyclic pure carbon ($C_5$) and five hydrogen atoms, which means dehydrogenation. At a step of n=6, there happens surprising separation and re-configuration as shown in green column in Figure 3. Two carbon atoms and two hydrogen atoms were separated far away. It should be noted that there occurs the creation of cyclic-$C_3H_2$ marked by red ellipse. Also we could notice a quasi-molecule of cyclic-$C_3H_3$ marked by blue. Cyclic-$C_3H_2$ is well known smallest PAH and current hot topic of advanced radio astronomy as referenced in introduction. In Figure4, calculated infrared spectrum of ($C_5H_5$)$^{n+}$ (n=0, 1,and 2) were illustrated.

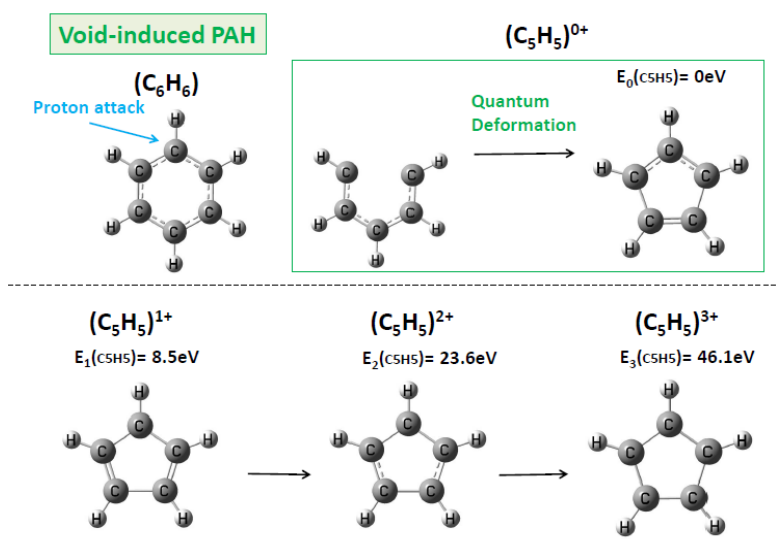

Figure 2, Defect induced ($C_5H_5$) and Photoionization from ($C_5H_5$)$^{0+}$ to ($C_5H_5$)$^{3+}$

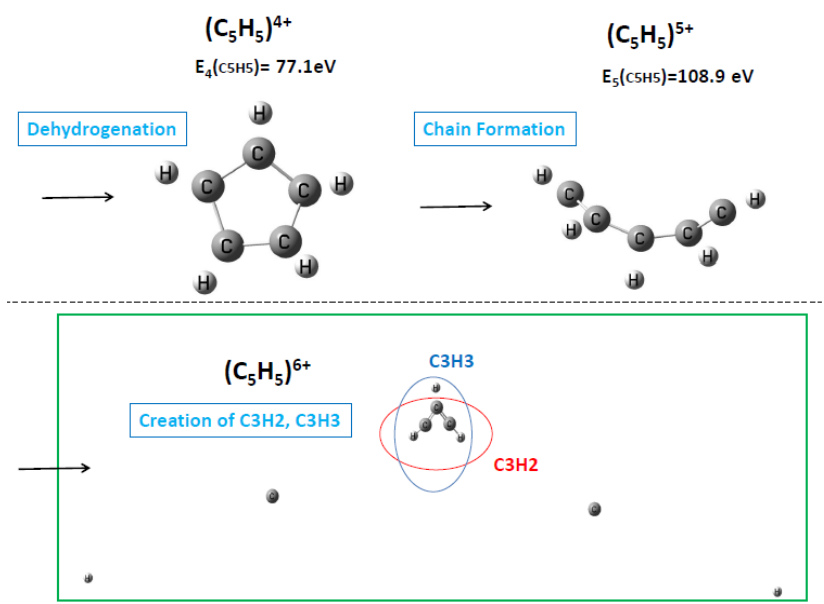

Figure 3, There occurs dehydrogenation at ($C_5H_5$)$^{4+}$, carbon chain formation at ($C_5H_5$)$^{5+}$, and the creation of cyclic-$C_3H_2$ (red circle in bottom) at ($C_5H_5$)$^{6+}$.

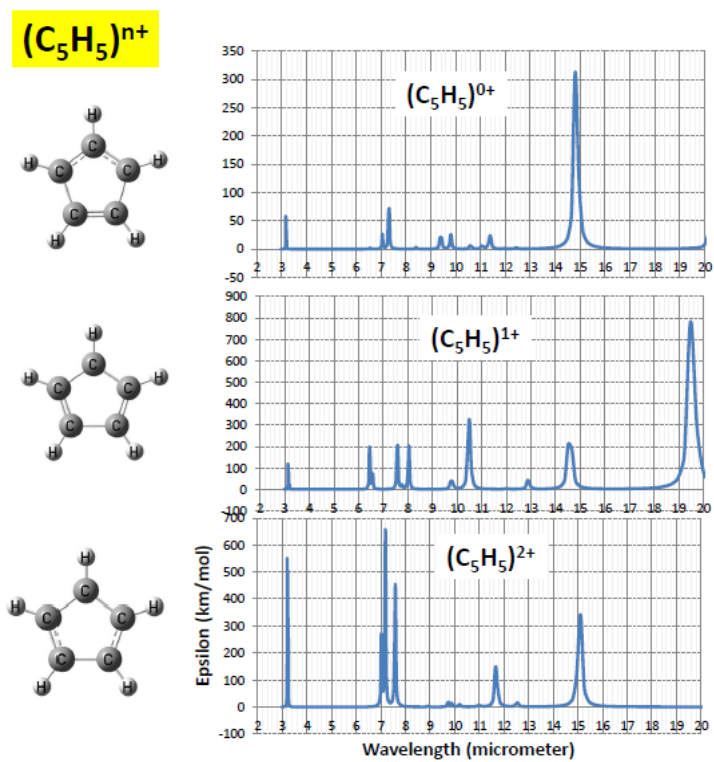

Figure 4, Calculated infrared spectrum of $(C_5H_5)^{n+}$

4, FROM CYCLIC-C3H2 TO CHAIN-C3

Deeper photoionization on cyclic-$C_3H_2$ was successively calculated as shown in Figure 5. Neutral and mono-cation molecules keep cyclic configuration as (cyclic-$C_3H_2$)$^{0+}$ and (cyclic-$C_3H_2$)$^{1+}$. At a step of di-cation, molecule shows drastic configuration change to aliphatic molecule (chain-$C_3H_2$)$^{2+}$. At the next ionization, chain-$C_3H_2$ was decomposed to chain-$C_3$ and two hydrogen atoms. Calculated infrared spectrum of $(C_3H_2)^{n+}$ (n=0,1, and 2) were shown in Figure 6.

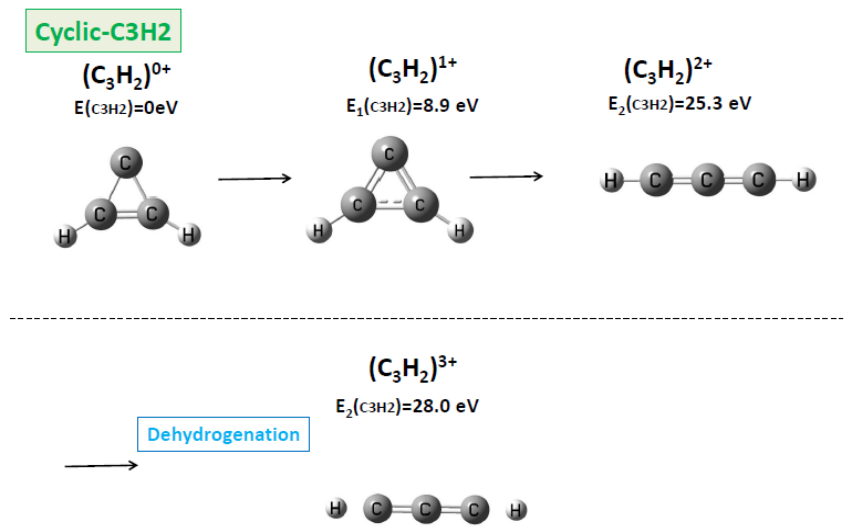

Figure 5, Deep photo-ionization from cyclic-$C_3H_2$ to aliphatic-$C_3H_2$, finally to carbon-chain $C_3$.

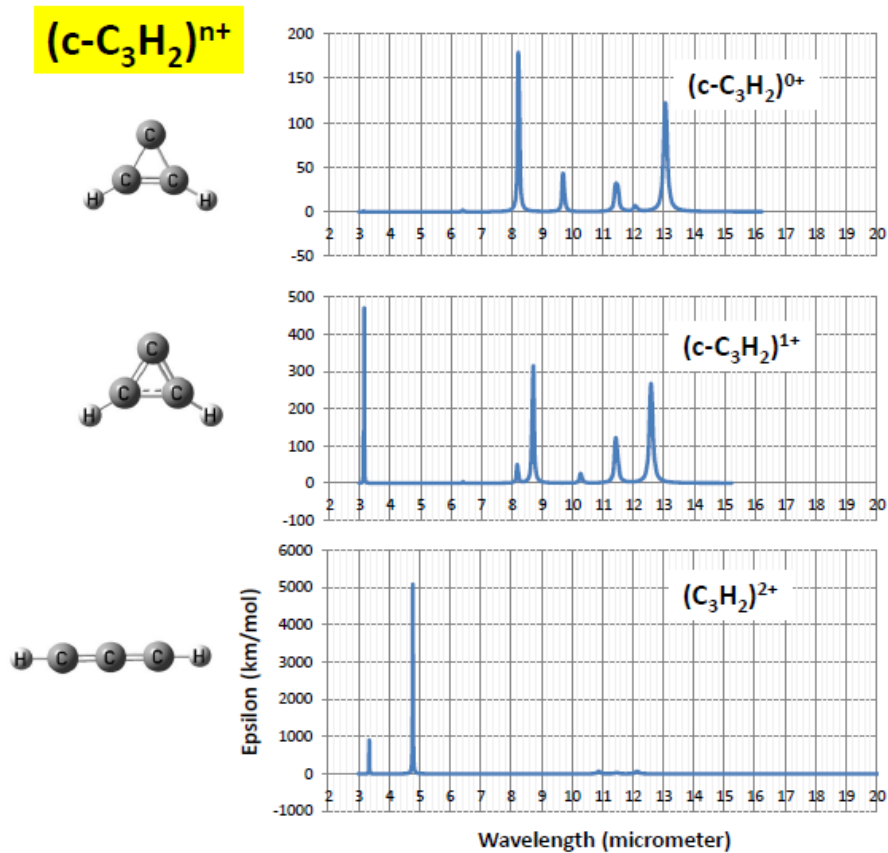

Figure 6, Calculated infrared spectrum of (cyclic-$C_3H_2$)$^{n+}$

## 5, INFRARED SPECTRUM OF (C5H5) AND (CYCLIC-C3H2)

We tried to identify observed infrared spectrum using these small molecules calculated one. In Herbig Ae young stars, most popular spectrum was ubiquitously observed Type-B (Ota 2017e) as illustrated in Figure 7. Typical star was HD34282. By a combination of (cyclic-$C_3H_2$)$^{1+}$ and ($C_5H_5$)$^{2+}$, major bands at 7.8, 8.6, 11.2, and 12.7 micrometer could be explained, but 6.2 micrometer band could not. Unfortunately, it is poor to reproduce band intensity ratio. In turn, as shown in Figure 8, single large molecule ($C_{23}H_{12}$)$^{2+}$ could reproduce the observed spectrum very well both wavelength and intensity ratio (Ota, 2017e). Small molecule could only partially reproduce observed spectrum.
 Another example is shown in Figure 9 in case of HD203024, which was classified as Type-D (Ota 2017e). Among many observed bands, four bands could be reproduced by a combination of ($C_5H_5$)$^{2+}$ and (cyclic-$C_3H_2$)$^{1+}$ marked by green dotted lines. Unfortunately, other many bands marked by red could not be reproduced. As noted in Figure 10, larger molecules combination by ($C_{12}H_8$)$^{n+}$ (n=0,1,and 2) could reproduce observed bands very well.

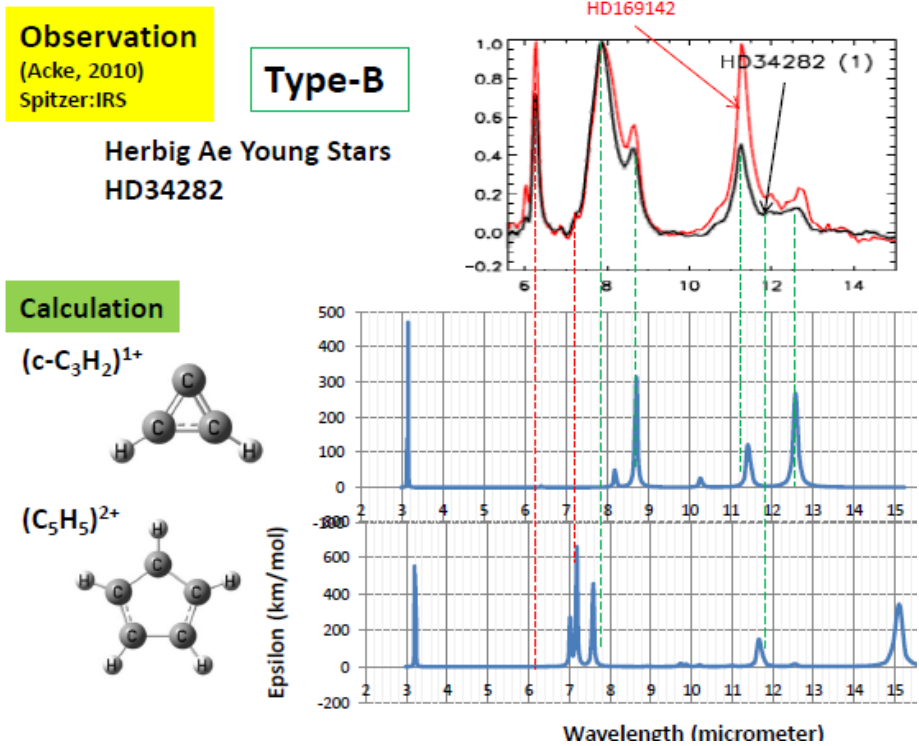

Figure 7, Identification of observed infrared spectrum of HD34282 by calculated combination of (cyclic-$C_3H_2$)$^{1+}$ and $(C_5H_5)^{2+}$.

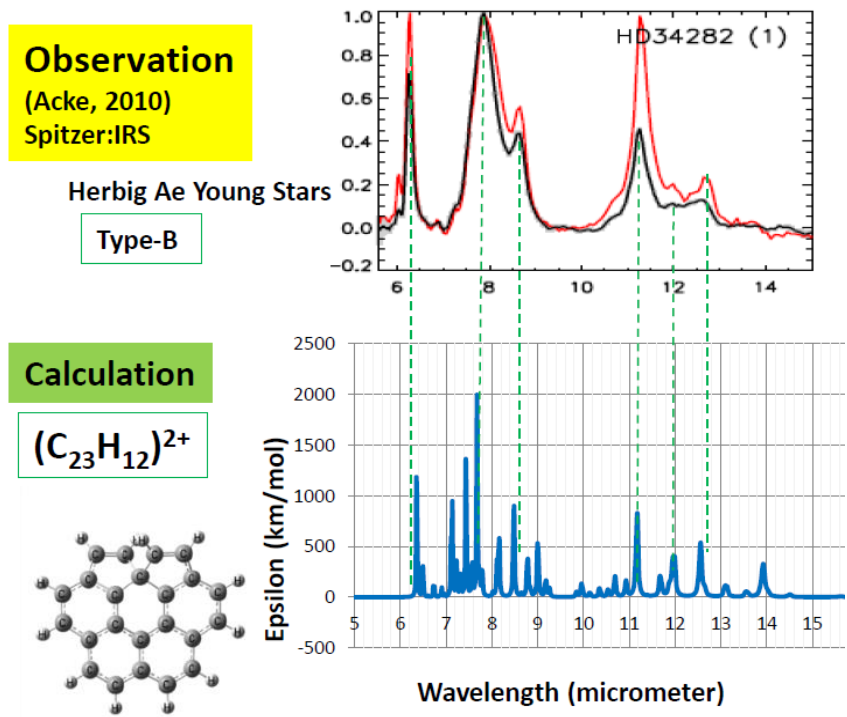

Figure 8, Identification of observed infrared spectrum of HD34282 by calculated one of $(C_{23}H_{12})^{2+}$.

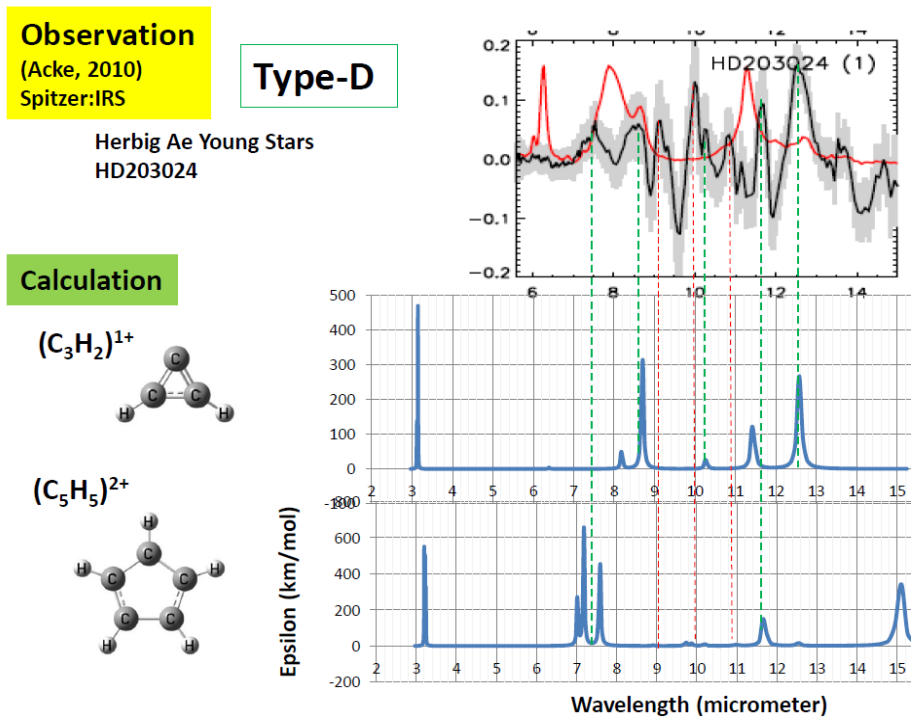

Figure 9, Identification of observed infrared spectrum of HD203024 by calculated one of (cyclic-$C_3H_2$)$^{1+}$ and $(C_5H_5)^{2+}$.

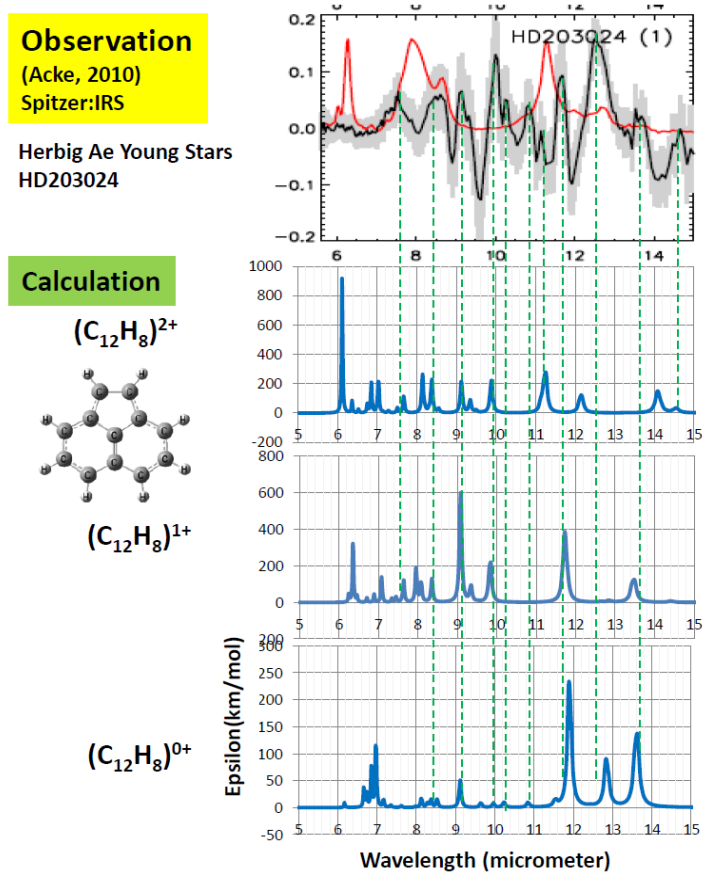

Figure 10, Identification of observed infrared spectrum of HD203024 by a mixture of calculated one of $(C_{12}H_8)^{n+}$ (n=0, 1, and 2)

## 6, RADIO ASTRONOMY OF (CYCLIC-C3H2)

Cyclic-$C_3H_2$ was observed and identified by radio astronomy. Radio astronomy can observe rotational mode of asymmetric molecule. Quantum-chemically calculated rotational constants A, B, and C were noted in Table 1. In case of small molecules like (cyclic-$C_3H_2$)$^{0+}$ and (cyclic-$C_3H_2$)$^{1+}$, A and B were 3GHz range and C was 1.6GHz range, which are suitable for radio frequency observation. Larger molecules, $(C_{23}H_{12})^{2+}$ and $(C_{12}H_8)^{2+}$, have rotational constants of 200~1500 MHz, which are difficult for current radio astronomy.

In 1985, by radio astronomy observation, Thaddeus et al. could report laboratory and astronomical identification of cyclic-$C_3H_2$ (Taddeus et al. 1985). Also in 1985, Matthews et al. opened that cyclic-$C_3H_2$ is ubiquitous in the Galaxy (Matthews et al. 1985). Detailed spacial distribution of cyclic-$C_3H_2$ in photo-dissociation region of the horsehead nebula was reported by Pety et al. (Pety et al. 2005). In 2013, using newly built radio observatory ALMA (Atacama large millimeter array), high resolution distribution of cyclic-$C_3H_2$ around a protoplanetary disk of young star HD163296 was reported by Qi et al. (Qi et al. 2013). Recently, surprising discovery was done by Sakai et al (Sakai et al. 2014) that cyclic-$C_3H_2$ is falling and rotating around a proto-stellar core (L1527) in the Taurus molecular cloud.

Combining above radio astronomy observation and our photo-ionization calculation, we can imagine a capable rout of cyclic-$C_3H_2$ in protoplanetary disk as illustrated in Figure 11. PAH molecules in interstellar cloud would be falling and rotating around the central young star. It should be noted that these PAHs will be photo-ionized by the central star and dissociated from larger to smaller one. Green line shows a falling and spiral curve of PAHs trajectory. By a centrifugal barrier of small molecules, hollow gap would be brought around the central star. Cross sectional image of protoplanetary disk is shown in Figure 12. Stable PAH like benzene ($C_6H_6$) may form the deep core of doughnuts like protoplanetary disk. By high speed proton attack, carbon void would be introduced, which will bring hydrocarbon pentagon ($C_5H_5$). By photon irradiation, ($C_5H_5$) would be reduced to cyclic-$C_3H_2$ and finally to pure carbon chain-$C_3$.

Astrochemical evolution step is summarized in Figure 13. It should be noted that these molecules could be identified in a natural way by quantum-chemical calculation introducing two astronomical phenomena, that is, void-induced quantum deformation and photo-ionization. Infrared astronomy observation is suitable for larger molecules and radio astronomy for smaller asymmetric molecules.

Table 1, Quantum-chemically calculated infrared spectrum, rotational constants and electric dipole moments

| | $(C_{23}H_{12})^{2+}$ | $(C_{12}H_8)^{2+}$ | $(C_3H_2)^{0+}$ | $(C_3H_2)^{1+}$ |
|---|---|---|---|---|
| Molecular Configuration | 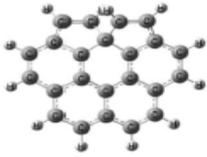 | 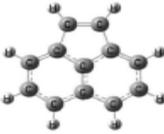 | 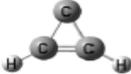 | 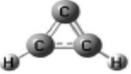 |
| Infrared Spectrum (Molecular Vibration) | 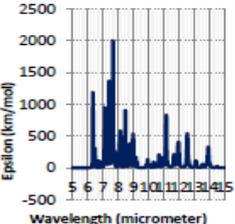 | 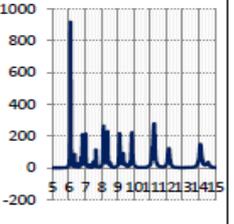 | 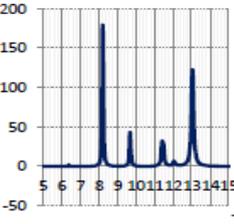 | 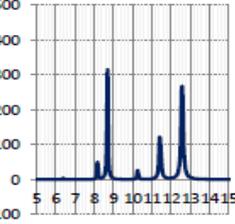 |
| Rotational Constants A(MHz) B C | 399.6 345.3 194.5 | 1558.5 1147.6 660.9 | 33029.5 31924.2 16233.7 | 38919.9 29381.4 16742.3 |
| Electric Dipole (Debye) | 1.20 | 1.67 | 3.20 | 1.34 |

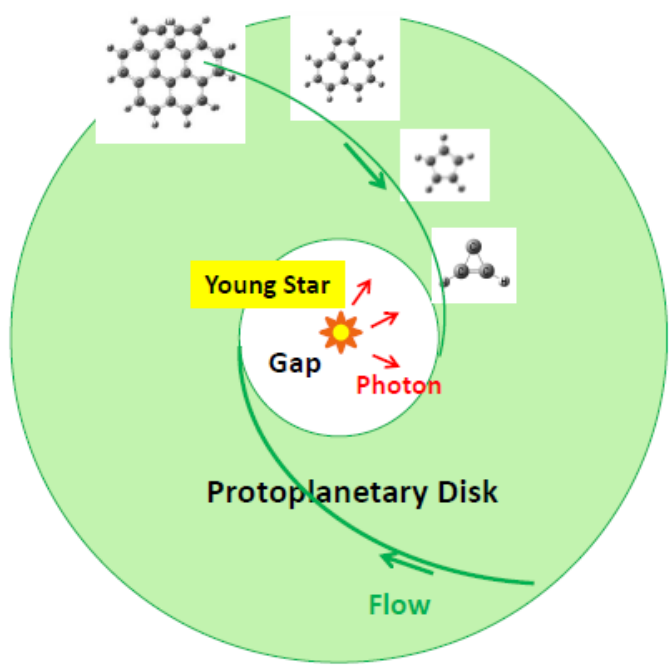

Figure 11, Trajectory of falling and rotating PAHs in a protoplanetary disk around a new born young star

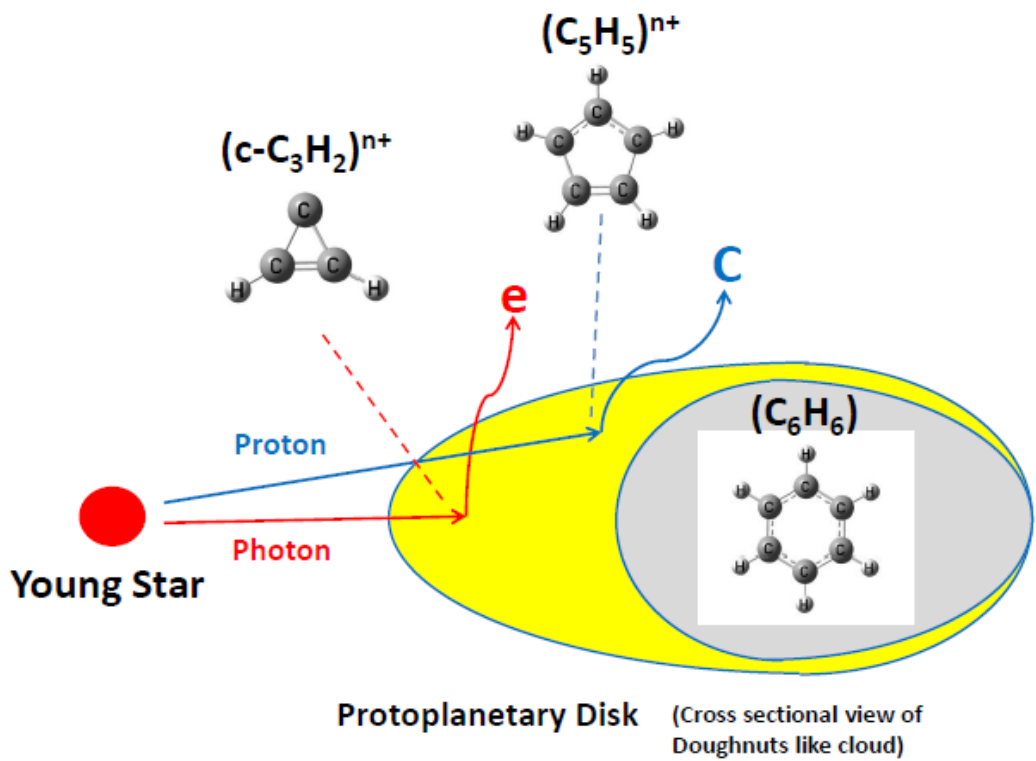

Figure 12, Cross sectional image of protoplanetary disk. There occurs chemical evolution of PAHs from ($C_6H_6$) to (cyclic-$C_3H_2$) by proton sputtering and photon irradiation from the new born star.

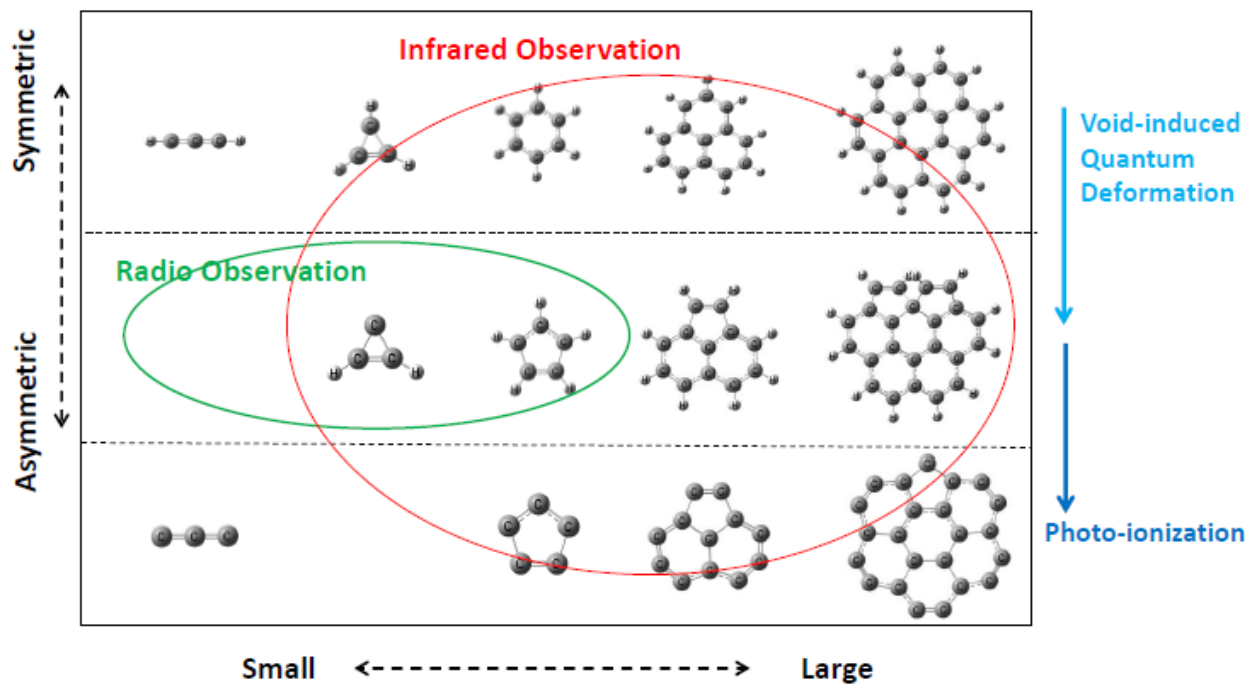

Figure 13, Evolution of PAHs and pure carbons by void-induced quantum deformation and photo-ionization. Green circled molecules could be observed by radio astronomy, whereas red marked ones by infrared astronomy.

## 7, CONCLUSION

Astronomical evolution mechanism of small size polycyclic aromatic hydrocarbon (PAH) was analyzed using the first principles quantum chemical calculation.

(1) Starting model molecule is benzene ($C_6H_6$), which would be transformed to ($C_5H_5$) due to carbon defect by high speed proton attack.
(2) In a protoplanetary disk around a new born young star like Herbig Ae star, molecules are illuminated by high energy photon from the central star and ionized to be cation ($C_5H_5$)$^{n+}$. Calculation shows that from n=0 to 3, molecule keeps original configuration. At ionization step n=4, there occurs separation to cyclic pure carbon ($C_5$) and five hydrogen atoms.
(3) At a step of n=6, there happens surprising creation of cyclic-$C_3H_2$, which is the smallest PAH. Cyclic-$C_3H_2$ had been usually observed and identified by radio astronomy observation.
(4) Deep photoionization was successively applied on cyclic-$C_3H_2$. Neutral and mono-cation $C_3H_2$ keep cyclic configuration as (cyclic-$C_3H_2$)$^{0+}$ and (cyclic-$C_3H_2$)$^{1+}$. At a step of di-cation, molecule shows a change to aliphatic chain ($C_3H_2$)$^{2+}$. At the next ionization, chain-$C_3H_2$ was reduced to chain-$C_3$ and two hydrogen atoms.
(5) Calculated infrared spectrum of those deep photo-ionized molecules was applied to observed spectrum of Herbig Ae stars. Observed typical infrared spectrum of HD34282 could be partially explained by a combination of calculated spectrum of (c-$C_3H_2$)$^{1+}$ and ($C_5H_5$)$^{2+}$. Meanwhile, excellent coincidence was obtained by applying a large molecule ($C_{23}H_{12}$)$^{2+}$. In case of HD203024, observed spectrum could be partially reproduced by those small molecules, but reproduced very well by a combination of ($C_{12}H_8$)$^{0+}$, ($C_{12}H_8$)$^{1+}$ and ($C_{12}H_8$)$^{2+}$.

It should be noted that these molecules could be identified in a natural way introducing two astronomical phenomena, that is, void-induced molecular deformation and deep photo-ionization. Infrared observation is suitable for larger molecules and radio astronomy for smaller asymmetric molecules.

Author profile
   Norio Ota PhD.
     Fellow and Honorable member, Magnetics Society of Japan
     Senior Professor, University of Tsukuba, Japan

   Specialty: Material Science, Magnetic and Optical data storage devices

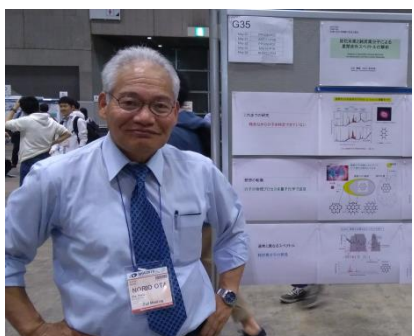

@ JpGU2018 Conference, May 2018

**Submit to arXiv.org.**

**October    xx, 2018    by Norio Ota**